\documentclass{article}
\usepackage{graphicx, subfigure}
\usepackage{amsmath}
\usepackage{enumerate}
\newtheorem{remark}{Remark}
\title{Phase-plane analysis of the timelike geodesics around a spherically symmetric static dilaton black hole}

\author{Paul Blaga\\Babe\c{s}-Bolyai University\\Faculty of Mathematics and Computer Science\\ Kog\u{a}lniceanu Street 1, 600410, Cluj-Napoca, Romania\\
	Email: pablaga@cs.ubbcluj.ro\\
	Cristina Blaga\\Babe\c{s}-Bolyai University\\Faculty of Mathematics and Computer Science\\
	Kog\u{a}lniceanu Street 1, 600410, Cluj-Napoca, Romania\\
	Email: cpblaga@math.ubbcluj.ro}

\begin{document}
\maketitle

\begin{abstract}
In this note we take a dynamical systems approach to the equations of motion of a free test particle moving around a spherically symmetric static dilaton black hole, written in the Einstein frame. The equations of motion are obtained using the Euler-Lagrange formalism. Using the first integrals of motion, we reach the conclusion that the free test particles are moving in a plane, named \emph{plane of motion}. In it we analyze the existence and nature of the equilibrium points and compare the behavior of free test particles near the equilibrium points using the dynamics systems approach. The study revealed that in the exact phase-plane exist distinct regions of motion, separated through a curve named \emph{separatrix}. In the end we  obtained a relation between the parameters describing the black hole and the free test particle that holds on a parabolic separatrix.	
\end{abstract}

\section{Introduction}

In classical theory of general relativity, the spacetime near a charged black hole is described using the Reissner-Nordstr\o m metric.

At the end of the eighties, especially after the publications of the classical monograph Green, Schwarz and Witten(1987)\cite{superstrings}, there was much interest in the investigation of black holes within string theory. The metric we use in this paper is a solution of the Einstein-Maxwell-dilaton equations, based on an action containing, beside gravity, a scalar field, called \emph{dilaton}, and the electromagnetic field, coupled to the graviton. More specifically, the action has the form
\begin{equation}\label{action}
S=\frac{1}{16\pi}\int d^4x\sqrt{-g}\left[R-2\left(\nabla\Phi\right)^2-e^{-2\Phi}F_{\mu\nu}F^{\mu\nu}\right]
\end{equation} 
where $g$ is the determinant of the metric, $R$ is the scalar curvature, $\Phi$ is the dilaton field, while $F_{\mu\nu}$ is the strength of the electromagnetic field.

The action~(\ref{action}) corresponds to the low-energy limit of the action for the heterotic string. It is written in the so-called \emph{Einstein frame}. In this approach, as one can see easily, the action is, simply, the action for pure gravity, with an energy-momentum defined by the electromagnetic fied and the dilaton (Einstein-Hilbert action). There is, also, an alternative approach, where the action is written in the so-called \emph{string frame}. In this approach, the metric from the action(\ref{action}) is replaced by another metric, conformally equivalent to the metric from the Einstein frame. In this paper, we shall work exclusively in the Einstein frame and postpone for another paper the discussion of the string frame. We notice, nevertheless, that the spacetimes obtained from the two actions are not isometric, they are just conformally equivalent. For more information about the two frames, see Casadio and Harms(1999)\cite{casadio} or the monograph Frolov and Novikov(1998)\cite{fronov}.

The first static spherically symetric black hole solution in dilaton gravity was found by Gibbons and Maeda(1988)~\cite{gm} and, three years later, independently, by Garfinkle, Horowitz and Strominger(1991)~\cite{ghs}. This solution, known as Gibbons--Maeda--Garfinkle--Horowitz--Strominger (GMGHS) black hole, was later reconstructed by Horowitz(1993)~\cite{hor}, through a Harrison-like transformation, starting from the Schwarz\-schild solution.  

In this article we perform a phase-plane analysis of the equations of motion of a free test particle around a GMGHS black hole. In section 2 we obtain the second order nonlinear equation. The phase-plane analysis for the corresponding two-dimensional system of first order equations of motion is performed in section 3. We obtain the equilibrium points and determine their nature. In the following section we represent several phase-plane diagrams and analyze the separatrix, curve which divide the phase-plane into distinct regions of motion. 

\section{Equations of motion}

The equations of motion of a free test particle moving around a GMGHS black hole are derived from the line element:  
\begin{equation}\label{metr}
ds^2=-\left(1-\frac{2M}{r}\right) dt^2 +\frac{d r^2}{\left(1-\frac{2M}{r} \right)} + r \left(r-\frac{Q^2}{M}\right) (d \theta^2 + \sin^2 \theta \, d \varphi^2)
\end{equation} 
where $Q$ is related to the electrical charge of the black hole and $M$ to its mass. For $Q^2 < 2 M^2$, the black hole has an events horizon. If $Q^2 = 2 M^2$, the solution describes a naked singularity. The latter case is known as  extremal GMGHS black hole. 

The Lagrangian corresponding to the line element~(\ref{metr}) is 
\begin{equation}\label{lagr}
2 \mathcal{L} = -\left(1-\frac{2M}{r}\right) \dot{t}^2 + \frac{\dot{r}^2}{\left(1-\frac{2M}{r} \right)} + r \left(r-\frac{Q^2}{M}\right) \left(\dot{\theta}^2  + \sin^2 \theta \,\dot{\varphi}^2 \right) 
\end{equation}
where dot means differentiation with respect to $\tau$ - an affine parameter along the geodesic. The parameter is chosen such that $2 \mathcal{L}=-1$ along a timelike geodesics, $2 \mathcal{L}=0$ along a null geodesics and $2 \mathcal{L}=1$ along a spacelike geodesics. 

The equations of motion of a free test particle are the Euler-Lagrange equations of the Lagrangian~(\ref{lagr}) (see Chandrasekhar(1983)\cite{cha}). The coordinates $t$ and $\varphi$ do not appear explicitly in~(\ref{lagr}), they are cyclic coordinates. Thus, one finds two integrals of motion. The first is derived from $\partial L/\partial t =0$, named the energy integral 
\begin{equation}\label{ien}
\left(1-\frac{2M}{r}\right) \dot{t} =  E
\end{equation}
where $E$ is a real constant -- the total energy of the particle. The second integral, obtained from $\partial L/\partial \varphi =0$, 
\begin{equation}\label{imc}
2 \, \sin^2\theta \, \cdot r \left(r-\frac{Q^2}{M}\right) \dot{\varphi} = \mbox{constant}
\end{equation}
is the angular momentum integral. 

The Euler-Lagrange equation for $\theta$ is 
\begin{equation}\label{ecteta}
\frac{d}{d \tau}\left[ r \left( r - \frac{Q^2}{M}  \right) \dot{\theta} \right] = r \left( r - \frac{Q^2}{M}  \right) \sin \theta \cos \theta \cdot \dot{\varphi}^2\, .
\end{equation}
If $\theta = \pi/2$, when $\dot{\theta}=0$, then from~(\ref{ecteta}) $\ddot{\theta} = 0$ and $\theta=\pi/2$ on the geodesic. And so, if at the beginning the free test particle is in the equatorial plane and $\dot{\theta}=0$, its motion is confined in the equatorial plane. The motion  is planar like in the Schwarzschild spacetime or in the Newtonian gravitational field.

If $\theta = \pi/2$ the angular momentum integral~(\ref{imc}) leads us to  
\begin{equation}\label{ecL}
r \left(r-\frac{Q^2}{M}\right) \dot{\varphi} = L
\end{equation} 
where the real constant $L$ is the angular momentum about an axis normal at the plane in which the motion took place.

Using the integrals of motion in the constancy of the Lagrangian we get a nonlinear first order differential equation in $r$
\begin{equation}\label{et}
\left( \frac{dr}{d \tau} \right)^2 + \left( 1 - \frac{2 M}{r} \right) \left(\frac{L^2}{r \left( r - \frac{Q^2}{M} \right)}  - \epsilon \right) = E^2
\end{equation}   
where $\epsilon=-1$ for timelike geodesics, $\epsilon=0$ for null geodesics and $\epsilon=+1$ for spacelike geodesics. In this article we are interested in the motion of free test particles around a GMGHS black hole, therefore we consider $\varepsilon=-1$. 

In analogy with the motion of a particle in the Newtonian gravitational field, the second term from the left-hand side of the relation (\ref{et}) is named \emph{effective potential}. For the timelike geodesics it is    
\begin{equation}\label{Veff}
V = \left( 1 - \frac{2 M}{r} \right) \left(\frac{L^2}{r \left( r - \frac{Q^2}{M} \right)}  + 1 \right) \,.
\end{equation} 

\section{Phase-plane analysis}

Denoting $x=r_S/r$, where $r_S=2M$ is the Schwarzschild radius, the relation~(\ref{et}) becomes
\begin{equation}\label{em}
\left(\frac{d x}{d \varphi} \right)^2 = 2 \sigma (1- b x)^2 \left(E^2 - \Lambda\right) - \left(1 - b x \right) \Lambda x^2
\end{equation}
where 
\begin{equation}\label{bs}
\sigma=\frac{1}{2} \left(\frac{r_{S}}{L}\right)^2, \quad b=\frac{Q^2}{2 M^2} \quad \mbox{and} \quad \Lambda=1-x
\end{equation}
We note that $\sigma>0$, $b \in \left[0,1 \right]$, $b=0$ is the Schwarzschild black hole, $b=1$ an extremal GMGHS black hole and outside the black hole $x \in [0,1]$.

We seek the solution $x=x(\varphi)$. Differentiating the equation~(\ref{em}), with respect to $\varphi$, we get 
\begin{equation}\label{oc}
\frac{d x}{d \varphi} = 0
\end{equation}
or
\begin{equation}\label{edm}
\frac{d^2 x}{d \varphi^2}= a_3 x^3 + a_2 x^2 +a_1 x+a_0 
\end{equation}
with
\begin{eqnarray}\label{ai}
a_3&=&-2 b \,,\qquad \qquad \qquad \qquad \quad a_2 = \frac{3}{2} \left(2 \sigma b^2 + b + 1 \right)\,,\nonumber\\ 
a_1&=&2 \sigma b^2 (E^2 - 1)-4\sigma b-1\,, \quad a_0= 2 \sigma b ( 1- E^2) + \sigma\,. 
\end{eqnarray}

The equation~(\ref{oc}) has the the particular solution $x=\mbox{constant}$ or $r$ is constant, \emph{i.e.} circular orbits, which were discussed in Blaga(2013)\cite{b13}. In this article, we study the equation~(\ref{edm}) using the dynamical systems approach (see Jordan and Smith(1999)\cite{js99} or Strogatz(1994)\cite{s94}).

\subsection{Equilibrium points}\label{par31}

We introduce the variable $y=dx/d\varphi$, to transform the second order, nonlinear, inhomogeneous differential equation~(\ref{edm}) into a first order differential system of equations
\begin{equation}\label{sis}
\begin{cases}
x'=y \\
y'=a_3 x^3 +a_2 x^2 +a_1x+a_0
\end{cases}
\end{equation}
where prime denotes the differentiation with respect to $\varphi$ and $a_i$, $i \in \{0,1,2,3\}$ are the coefficients of equation~(\ref{edm}), given by~(\ref{ai}). 

The equilibrium points of the system~(\ref{sis}) are given by $x'=y'=0$. To find them, we solve simultaneously the equations for $x$ and $y$. The equation $x'=0$ has the solution $y^{\star}=0$. If we write $E^2$ in terms of $x$, $\sigma$ and $b$ from~(\ref{em}) for $dx/d \varphi=x'=0$, we get
\begin{equation}\label{E2}
E^2=(1-x)\left[ 1+ \frac{x^2}{2 \sigma (1- bx)}\right]
\end{equation}
and replacing~(\ref{E2}) in the second equation of the system~(\ref{sis}), we obtain 
\begin{equation}\label{ceq}
y'=- b x^3 + \left(\sigma b^2 +\frac{1}{2} b + \frac{3}{2} \right) x^2 - \left(2 \sigma b +1\right) x + \sigma \,.
\end{equation}

From $y'=0$, we get a cubic equation in $x$, with coefficients depending on $b$ and $\sigma$. The leading coefficient is equal to $b$. If $b=0$,~(\ref{ceq}) becomes a quadratic equation in $x$. The case $b=0$, represents a Schwarzschild black hole and it was analyzed, by using dynamical systems, by Dean(1999)\cite{db99}.  

If $b \neq 0$, the abscissae of the equilibrium points are the roots of the cubic equation $y'=0$. Using Cardan's formula (see for example Kurosh(1980)\cite{k80}) these are  
\begin{equation}
x^{\star}_1=d + u + v\,,\,x^{\star}_2=d - \frac{u+v}{2} + \frac{u-v}{2} \sqrt{-3}\,,\,x^{\star}_3=d - \frac{u+v}{2} - \frac{u-v}{2} \sqrt{-3}
\end{equation}
where
\begin{equation}
d=\frac{2\sigma b^2+b+3}{6b} \,,\, u=\sqrt[3]{-\frac{q}{2}+\sqrt{\Delta}} \,,\, v=\sqrt[3]{-\frac{q}{2}-\sqrt{\Delta}} \,,\,\Delta=\frac{q^2}{4} + \frac{p^3}{27}
\end{equation}
and 
\begin{equation}
p=-\frac{w^2}{12b^2} \,,\,q=-\frac{w^3+54(1-b)}{108b^3}\,,\,w=2\sigma b^2+b-3\,.
\end{equation}
The nature of roots of the cubic equation $y'=0$ depends on the sign of the discriminant $\Delta$, which, after some algebra, becomes
\begin{equation}\label{Delta}
\Delta=\frac{(1-b)[8b^4\sigma^3+12b^2(b-3)\sigma^2+6(b-3)^2\sigma+b-9]}{432b^4}\,.
\end{equation} 
If $\Delta>0$ the equation has one real and two conjugate complex roots, if $\Delta=0$ it has three real roots, at least two equal and if $\Delta<0$ it has three distinct, real roots. 

In our analysis, $0 < b \leq 1$ and $\sigma>0$, therefore the sign of discriminant is the sign of second factor from the numerator. We consider it as a fourth order polynomial in $b$, with real coefficients, depending on the parameter $\sigma$, denoted with $h(b, \sigma)$. We can write it like   
\begin{equation}
h(b, \sigma) = 8 \sigma^3 b^4 + 12 \sigma^2 b^3 - 6 \sigma (6 \sigma -1) b^2+(1-36 \sigma) b + 9 (6 \sigma -1)\,.
\end{equation}
with $h(0, \sigma)=9(6 \sigma -1)$ and $h(1, \sigma) = 8 (\sigma -1) ^3$. Using the Sturm method (see for example Kurosh(1980)\cite{k80}) we establish the number of roots of this polynomial for  $b \in [0,1]$ and  $\sigma>0$. In the Sturm's sequence, the first term is $h(b, \sigma)$. The second term is the first derivative of $h(b, \sigma)$ with respect to $b$, 
\begin{equation}
h'(b, \sigma)=32 \sigma^3 b^3 + 36 \sigma^2 b^2 - 12 \sigma (6 \sigma -1) b+(1-36 \sigma)\,.
\end{equation}
The third term is the reminder after dividing $h$ by $h'$, with reversed sign,
\begin{equation}
h_2(b, \sigma) = \frac{3}{32 \sigma} [192 b^2 \sigma^3 +4 b^2 \sigma^2 + 4 b \sigma (54 b +1) - 576 \sigma^2 +60 \sigma +1]=\frac{3}{32 \sigma} h_{2s}\,,
\end{equation}
followed by the reminder after dividing $h'$ by $h_{2s}$, with reversed sign,
\begin{equation}
h_3(b, \sigma) = \frac{576 \sigma^2}{(48 \sigma + 1)^2} [-4 \sigma (24 \sigma+1) b + 144 \sigma +1]=\frac{576 \sigma^2}{(48 \sigma + 1)^2} h_{3s}\,
\end{equation}
and the reminder after dividing $h_{2s}$ by $h_{3s}$, with reversed sign, 
\begin{equation}
h_4(b, \sigma) = \frac{9 (48 \sigma + 1)^2}{4(24 \sigma + 1)^2} (64 \sigma^2 -88 \sigma -1)\,.
\end{equation}
During division process we have multiplied and divided by arbitrary positive quantities, because only the sign of the reminder matters in the Sturm method. 

The change of sign for these polynomials if $b \in \{0,1\}$ and $\sigma \geq 0$ is given in Table~\ref{tab1} and Tabel~\ref{tab2}. If $b$ goes from $0$ to $1$, the Sturm sequence loses one change in sign if $\sigma \in [1/6,1]$, therefore we conclude that $h(b,\sigma)=0$ has one root in the interval $b\in[0,1]$ for $\sigma \in [1/6,1]$. And so, if  $b \in [0,1]$ and $\sigma>0$ then    
\begin{enumerate}[I.]
	\item for $0<\sigma \leq 1/6$, then $\forall b \in [0,1], h(\sigma,b) < 0$,  
	\item for $1/6<\sigma<1$, $\exists \bar{b} \in [0,1]$ so that $h(\sigma,\bar{b})=0$ and
	\begin{enumerate}[a.]
		\item if $b \in [0,\bar{b})$ then  $h(\sigma,b) > 0$ or
		\item if $b = \bar{b}$ then  $h(\sigma,\bar{b}) = 0$ or
		\item if $b \in (\bar{b},1]$ then  $h(\sigma,b) < 0$,
	\end{enumerate}
	\item for $\sigma \ge 1$ then $h(\sigma,b) > 0$. 	
\end{enumerate} 

\begin{table}
	\centering
	\caption{The number of sign changes for $b=0$ and $\sigma \geq 0$}\label{tab1}
	{\footnotesize
		\begin{tabular}{|c|cccccccccc|}
			\hline
			$\sigma$&0&&1/36&&0.12&&1/6&&1.39&\\
			\hline
			$h$&-&-&-&-&-&-&0&+&+&+\\ \hline
			$h'$&+&+&0&-&-&-&-&-&-&-\\ \hline
			$h_2$&+&+&+&+&0&-&-&-&-&-\\ \hline
			$h_3$&0&+&+&+&+&+&+&+&+&+\\ \hline 
			$h_4$&-&-&-&-&-&-&-&-&0&+\\ \hline
			Sign&&&&&&&&&&\\
			changes&&2&&2&&2&&3&&2\\ 
			\hline
		\end{tabular}
	}
\end{table} 

\begin{table}
	\centering
	\caption{The number of sign changes for $b=1$ and $\sigma \geq 0$}\label{tab2}
	{\footnotesize
		\begin{tabular}{|c|cccccccccccccccc|}
			\hline
			$\sigma$&0&&0.04&&0.22&&1&&1.39&&1.47&&1.59&&1.65&\\
			\hline
			$h$&-&-&-&-&-&-&0&+&+&+&+&+&+&+&+&+\\ \hline
			$h'$&+&+&0&-&-&-&-&-&-&-&-&-&0&+&+&+\\ \hline
			$h_2$&+&+&+&+&0&-&-&-&-&-&-&-&-&-&0&+\\ \hline
			$h_3$&0&+&+&+&+&+&+&+&+&+&0&-&-&-&-&-\\ \hline 
			$h_4$&-&-&-&-&-&-&-&-&0&+&+&+&+&+&+&+\\ \hline
			Sign&&&&&&&&&&&&&&&&\\
			changes&&2&&2&&2&&3&&2&&2&&2&&2\\
			\hline
		\end{tabular}
	}
\end{table} 

We recall that the number of roots of the equation $y'=0$, depends on the sign of the discriminant $\Delta$, which is the same with the sign of $h(\sigma,b)$. The roots of the cubic equation $y'=0$ are the abscissae of the equilibrium points of the system~(\ref{sisf}). The ordinates vanish for all the equilibrium points. Therefore, there are: three equilibrium points in the cases I and II.c, two equilibrium points for II.b and one equilibrium point for II.a and III. In the case II.b, one root of the equation $\Delta=0$ is a double root and the corresponding equilibrium point is a cusp.

\begin{remark}
	The equilibrium points of the system~(\ref{sis}) are the extremal points of the effective potential~(\ref{Veff}).
\end{remark}

The function~(\ref{Veff}) has always a minimum point, inside the events horizon~(see Blaga(2013)\cite{b13}). If $V$ admits three extremal points, the other two are outside the events horizon, one is a minimum and the other a maximum. If there are two extremal points, the point outside events horizon is an inflection point for the potential.

\subsection{Linear stability analysis}

A classification of the equilibrium points could be obtained using the linear stability analysis. First, we expand in Taylor series the right hand side of the equations
\begin{equation}\label{sisf}
\begin{cases}
x'=y  \\
y'=- b x^3 + \left(\sigma b^2 +\frac{1}{2} b + \frac{3}{2} \right) x^2 - \left(2 \sigma b +1\right) x + \sigma 
\end{cases}
\end{equation}
about the fixed points, in small parameters $\delta x = x-x^{\star}$ and $\delta y = y-y^{\star}$. Dropping the second order terms, we get the first order linear equations near the equilibrium point $(x^{\star},y^{\star})$    
\begin{equation}\label{sislin}
\begin{cases}
\delta x'= \delta y  \\
\delta y'=\left[-3b{x^{\star}}^2+(2\sigma b^2 +b +3)x^{\star}-2\sigma b -1\right]\delta x.
\end{cases}
\end{equation}
The general solution of the system~(\ref{sislin}) is an exponential. The nature of equilibrium points of the linearized system~(\ref{sislin}), depends on the eigenvalues of matrix
\begin{equation}
A=
\begin{pmatrix}
0 & 1\\
-(1-bx^{\star})(2\sigma b+1-3x^{\star})&0
\end{pmatrix}\,,
\end{equation}
which are       
\begin{equation}
	\lambda_{1,2}=\frac{1}{2}\left( \tau \pm \sqrt{\tau^2-4 D} \right)
\end{equation}
where $\tau$ is the trace of the matrix $A$ and $D$ its determinant. The trace of the matrix $\tau=0$ and its determinant is 
\begin{equation}\label{D}
	D=(1-bx^{\star})(2\sigma b +1 - 3 x^{\star})\,.
\end{equation}
Therefore, the eigenvalues of matrix $A$ are 
\begin{equation}
	\lambda_{1,2}=\pm \sqrt{-(1-bx^{\star})(2\sigma b +1 - 3 x^{\star})}\,.
\end{equation}

If $D<0$, the eigenvalues are real numbers, with opposite sign and the equilibrium point is an unstable saddle. Its stability is not affected by small nonlinear terms. If $D>0$, the eigenvalues are purely imaginary and the fixed point is a center. The orbits around it are ellipses. Thus, we conclude that, among the solutions of nonlinear system~(\ref{sisf}), one can find precessing ellipses. 

Having in mind that if the equilibrium point is outside the events horizon, its abscissa $x^{\star}<1$, and the sign of $D$ is the sign of second factor from~(\ref{D}). It is easy to check that if $x^{\star}>1$, for $b \in [0,1]$ and $\sigma>0$, both factors from~(\ref{D}) are negative and $D>0$. Therefore, the equilibrium point, situated inside the events horizon, is a center.   

In the end, let us recall that if the dynamical system is conservative, \emph{i.e} admits a function that is constant on trajectories, then the equilibrium points coincide with the extremal points of that function (see Jordan and Smith (1999)\cite{js99}). In our case the constant function is the effective potential, therefore the equilibrium points are minima or maxima of $V$. A minimum point of the potential is a center, a maximum point is a saddle point. An inflection point is a cusp. Based on these observations and the study of effective potential $V$ done in Blaga(2013)\cite{b13}, we can conclude that if the system~(\ref{sis}) admits three equilibrium points, two are centers and one is a saddle point. If there are only two equilibrium points, one is a cusp and the other is a center. Outside the events horizon is the cusp, obtained through the merging of a center and the saddle point.         

\subsection{Phase plane diagram}

Using the linear stability analysis we obtain information about the behavior of the solution near the equilibrium points. The global features of the orbits around the black hole are revealed by the phase plane diagram for the nonlinear system~(\ref{sis}). The phase paths satisfy the separable differential equation 
\begin{equation}
\frac{d y}{d x}= \frac{a_3 x^3 + a_2 x^2 + a_1 x + a_0}{y}\,,
\end{equation}
where $a_i$, $i=\overline{0,3}$ are given by~(\ref{ai}), which give us, through integration, the level curves 
\begin{equation}\label{lc}
y^2 = \frac{1}{2} a_3 x^4 + \frac{2}{3} a_2 x^3 + a_1 x^2 + 2 a_0 x + \mathcal{C}\,,
\end{equation}
where $\mathcal{C}=2 \sigma (E^2-1)$ according to the equation~(\ref{em}). A complete study of the level curves for $b=0$, a Schwarzschild black hole, was done by Dean(1999)\cite{db99}.

\begin{figure}[ht!]
	\begin{center}
		\subfigure[$\sigma=1/7$ and $b=1/2$]{%
			\label{fig:first}
			\includegraphics[width=0.5\textwidth]{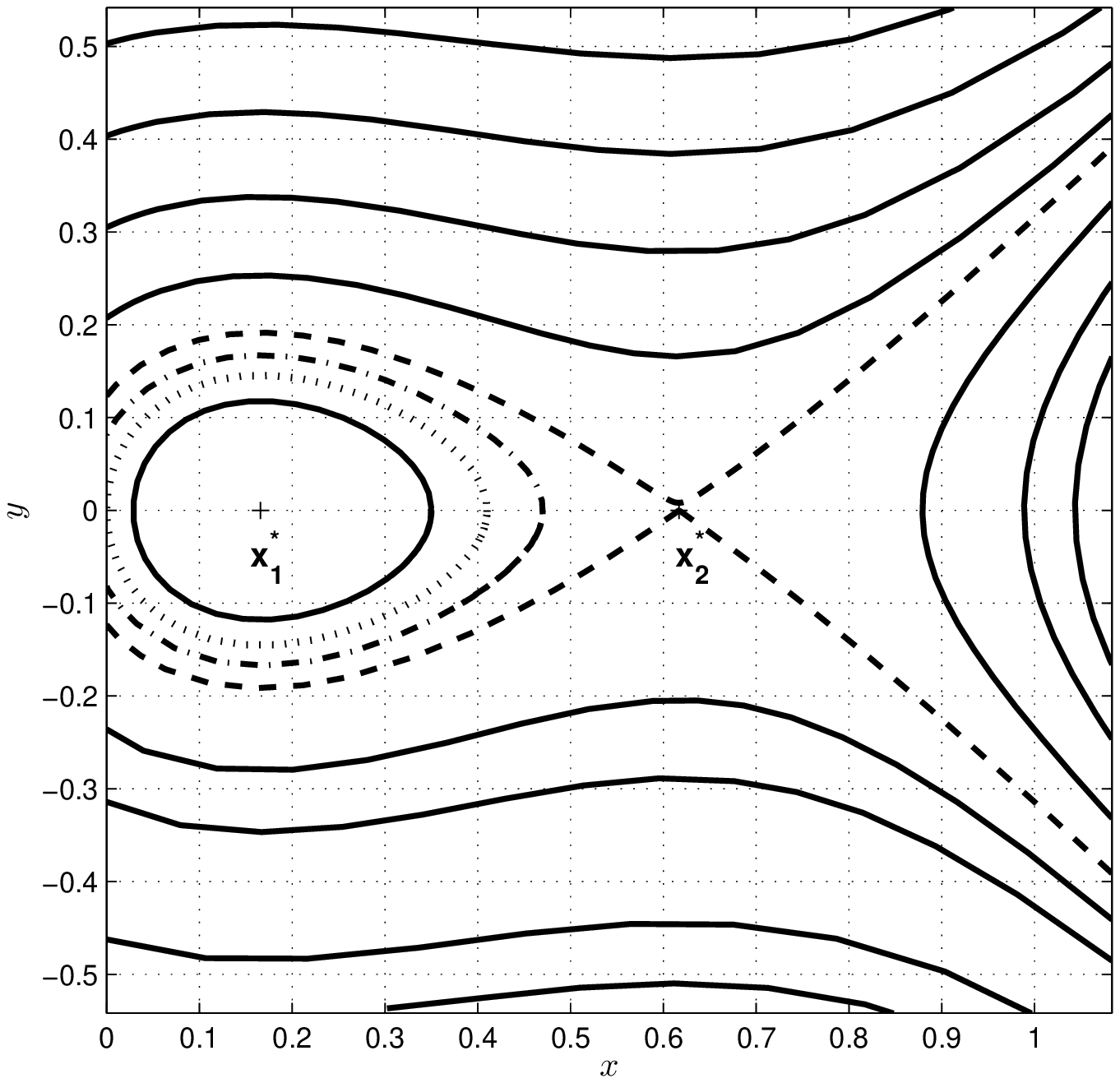}
		}%
		\subfigure[$\sigma=1/5$ and $b=1/5$]{%
			\label{fig:second}
			\includegraphics[width=0.5\textwidth]{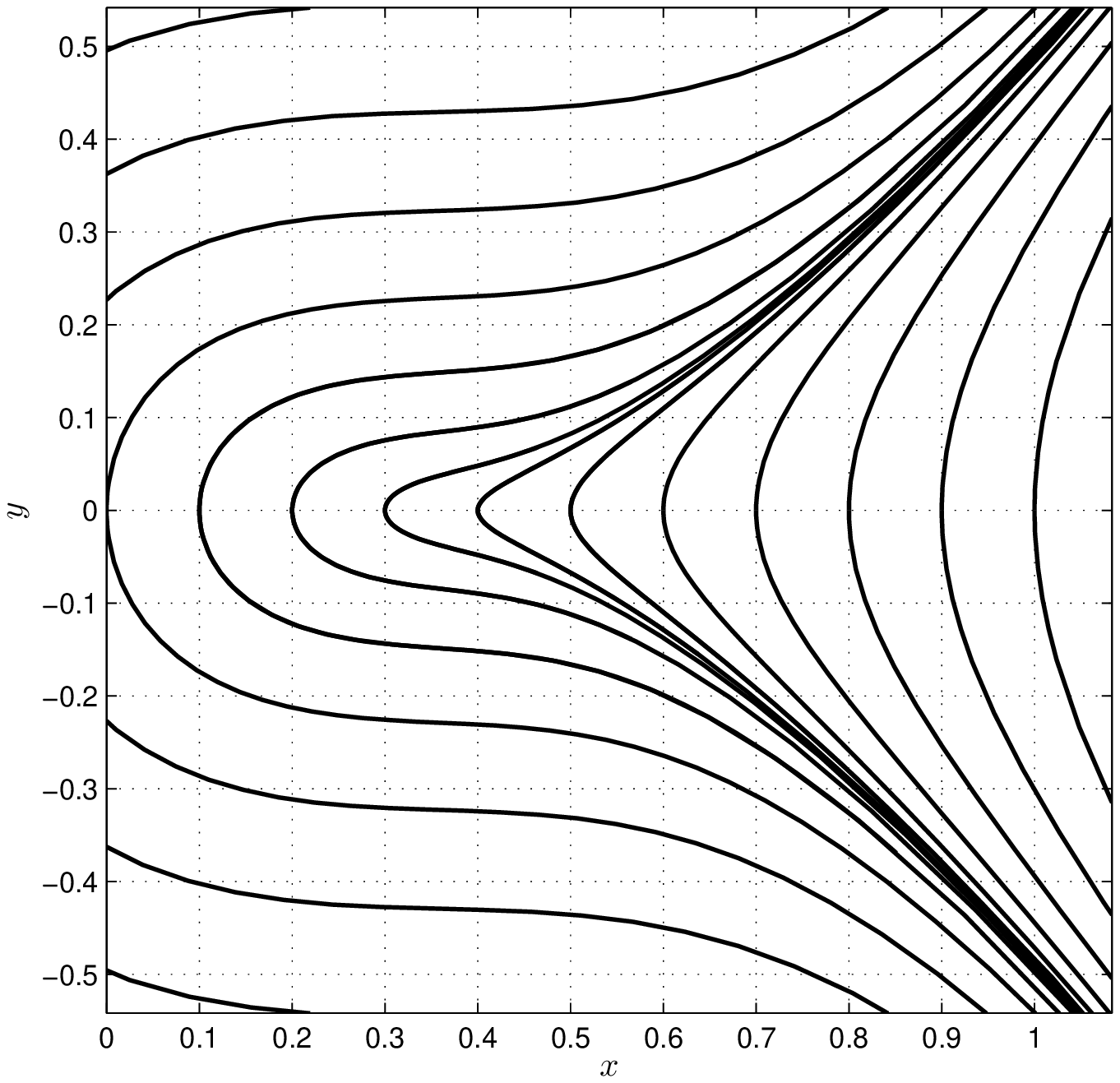}
		}\\ 
		\subfigure[$\sigma=1/5$ and $b=\bar{b}=0.289$]{%
			\label{fig:third}
			\includegraphics[width=0.5\textwidth]{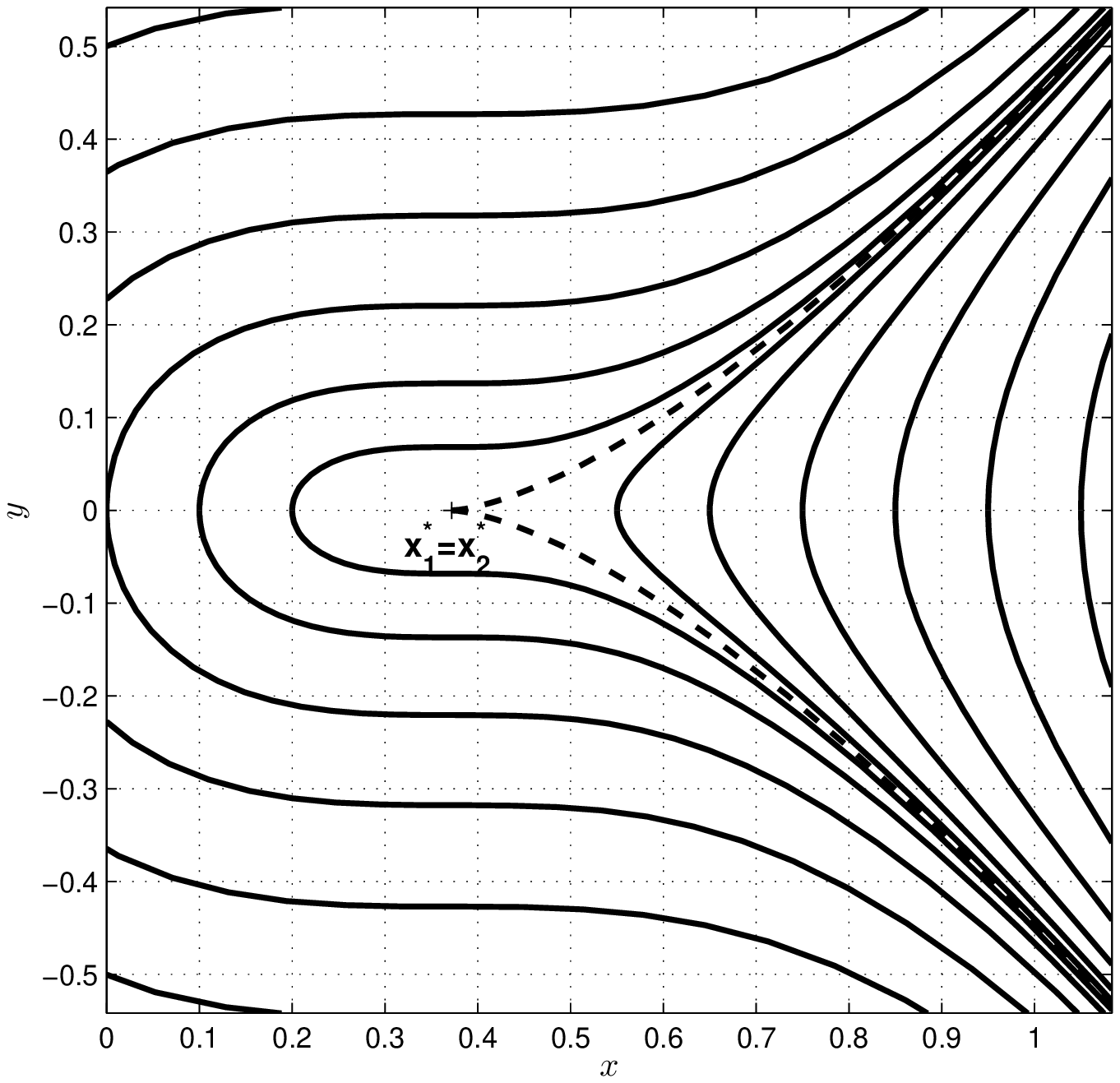}
		}%
		\subfigure[$\sigma=1/5$ and $b=1/2$]{%
			\label{fig:fourth}
			\includegraphics[width=0.5\textwidth]{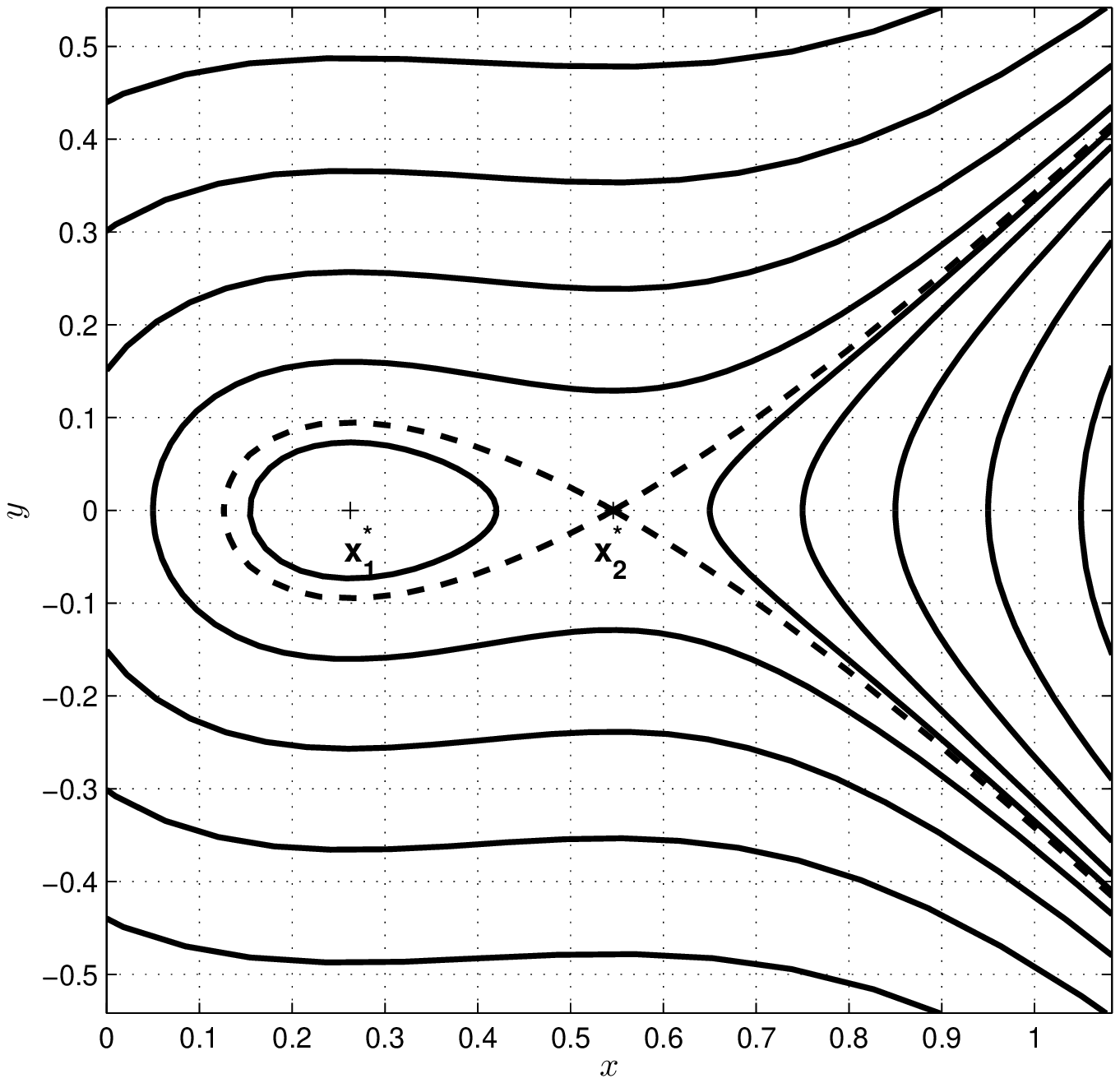}
		}
	\end{center}
	\caption{%
		Phase portrait for different values of $\sigma$ and $b$. We have represented with $+$ the equilibrium points situated outside the horizon, with dashed line the separatrix and, in (a) and (d), inside it, with solid line an elliptic orbit and in (d) with dotted line the parabolic orbit and with dashdot line a hyperbolic orbit.
	}%
	\label{f1}
\end{figure}

In figure~\ref{f1} we have represented the phase-plane diagram for different values of the parameters $\sigma$ and $b$. In the first plot from figure~\ref{f1} we sketched the exact phase plane diagram for a black hole with $\sigma=1/7$ and $b=1/2$, for different values of energy, outside the events horizon. In this case, representative for the case I, $\sigma \leq 1/6$, from section~\ref{par31}, there are always three equilibrium points. The homoclinic path which joins the saddle point $x_2^{\star}$ to itself, named \emph{separatrix}, is plotted with dashed line in figure~\ref{f1}. This phase path gives a graphic representation of the relation between black hole, angular momentum and energy on the unstable circular orbit and separates distinct regions in the phase plane. Inside it, is located the equilibrium point $x_1^{\star}$ -- a node for the linearized system~(\ref{sislin}). The third equilibrium point, $x_3^{\star}$ (a node), is inside the events horizon. 

The rest of the plots from figure~\ref{f1} contain examples for the case II: $1/6 < \sigma < 1$, from section~\ref{par31}. We choose $\sigma = 1/5$, for which $\bar{b}=0.289$, therefore we have considered $b \in \{1/5,0.289, 1/2\}$. For $b=1/5$, figure~\ref{f1}b there is only an equilibrium point, inside events horizon. In figure~\ref{f1}c, $b=\bar{b}=0.289$, and the equilibrium points outside events horizon coincides, the point $x_1^{\star}=x_2^{\star}$ being a cusp. For $b=1/2$, figure~\ref{f1}d, there are three equilibrium points, two of them outside the horizon. If $\sigma \geq 1$, case III in section~\ref{par31}, the phase portrait looks like that from figure~\ref{f1}b, because in that case, the only equilibrium point of the system~(\ref{sisf}) is inside the events horizon.       

We rewrite the terms from the right hand side of the level curve~(\ref{lc}) like 
\begin{equation}\label{cnr}
y^2= 2 \sigma (1-b x)^2 (E^2-1+x) - (1-b x)(1-x)x^2
\end{equation} 
to analyze the intersection of the separatrix with the $Oy$ axis. If $x=0$ in~(\ref{cnr}), we get $y^2= 2 \sigma (E^2-1)$ and recalling that $\sigma>0$, we obtain that the separatrix intersect the $Oy$ axis if and only if $E^2 \geq 1$, $E$ being the energy of the test particle for the unstable circular orbit.

If $x=0$, $r$ goes to infinity, thus if the separatrix cuts the $Oy$ axis, the motion is unbounded. If $E^2=1$, the orbit is a parabola, and if $E^2>1$ it is a segment of hyperbola. These results are in good agreement with the numerical investigation performed by~Olivares and Villanueva(2013)\cite{ov} and Blaga(2015)\cite{b15}.

If the separatrix is a parabolic orbit, $E^2=1$, the level curve~(\ref{cnr})  becomes
\begin{equation}\label{ps}
y^2 = x (1-b x) [ x^2-(2 \sigma b +1) x + 2 \sigma ]\,,
\end{equation}   
and it should go through the points $(0,0)$ and $(x_2^{\star},0)$. Thus, among the roots of the equation obtained substituting in~(\ref{ps}) $y=0$, we should find $x=0$ and $x=x_2^{\star}$. But $0<b\leq 1$ and $\sigma>0$, therefore $x_2^{\star}$ should be a solution of the third factor in~(\ref{ps}), the following quadratic equation in $x$
\begin{equation}\label{ec2}
x^2-(2 \sigma b +1) x + 2 \sigma=0 \,.
\end{equation}   
Let us remember that $x_2^{\star}$ is a root of the cubic equation~(\ref{ceq}). Two polynomials with arbitrary coefficients have a common root if and only if their resultant is zero (see for example Kurosh(1980)\cite{k80}). The resultant the quadratic and cubic polynomials on hand, $R$, is 
\begin{equation}\label{res}
R=\frac{1}{2}\,\sigma \left( -1 + b \right)  \left( 4 \sigma^2 b^2 + 4 \sigma b + 1 - 8 \sigma \right)\,, 
\end{equation}     
and, having in mind that $\sigma >0$ and $b \in (0,1]$ we obtain that it is equal with zero if $b=1$ or if 
\begin{equation}\label{dis}
4 \sigma^2 b^2 + 4 \sigma b + 1 - 8 \sigma =0\,.
\end{equation} 
On one hand, we note that the expression from the left side of the relation~(\ref{dis}) is the discriminant of the quadratic equation~(\ref{ec2}). If it is zero,~(\ref{ec2}) admits a double root, $x=(2 \sigma b +1 )/2$.

On the other hand, having in mind the meaning of the parameters $b$ and $\sigma$, introduced by the relation~(\ref{bs}), for a black hole, specified through $b$, and a free particle, defined through $\sigma$, if~(\ref{dis}) holds, the particle moves on a parabolic separatrix. Let us mention that if $b=0$, from~(\ref{dis}), we get $\sigma =1/8$ on the parabolic separatrix, result obtained by Dean(1999)\cite{db99}.   

\section{Conclusions}

Using the dynamical systems approach we studied the equations of motions around a spherically symmetric static dilaton black hole. Compared to the classical methods, like expression of the solution in terms of elliptic $\wp$-Weierstrass function or numerical integration, the main features of the motion are revealed more easily, using knowledge of dynamical systems and algebra. In the exact phase-plane there are distinct regions of motion, separated by the separatrix. Analyzing it we have obtained a relation between the parameters describing the black hole and the particle, which holds on the parabolic separatrix, the border between the bounded and unbounded motion.      


\end{document}